\documentclass[preprint2]{aastex}
\usepackage{epsfig}

\newcommand{\be}{\begin{equation}}
\newcommand{\ee}{\end{equation}}

\begin{document}

\title{
Dark energy and the mass of the Local Group
}

\author{A. D. Chernin \altaffilmark{1,2} \altaffiltext{1}{Sternberg 
Astronomical Institute, Moscow
University, Moscow, 119899, Russia}
P. Teerikorpi
\altaffilmark{2} \altaffiltext{2}{Tuorla Observatory, Department of 
Physics and Astronomy, University of Turku, SF-21500 Piikki\"o,
Finland}
M. J. Valtonen
\altaffilmark{2}
G. G. Byrd
\altaffilmark{3} \altaffiltext{4}{University of Alabama, Tuscaloosa, AL 
35487-0324, USA }
V. P. Dolgachev
\altaffilmark{1}
and L. M. Domozhilova
\altaffilmark{1}
}

\date{Received / Accepted}

\begin{abstract}

Dark energy must be taken into account to estimate more reliably the 
amount of dark matter and how it is
distributed in the local universe. For systems several Mpc across like 
the Local Group, we
introduce three self-consistent independent mass estimators. These 
account for the antigravity
effect of dark energy treated as Einstein's cosmological
constant $\Lambda$. 
The first is a modified Kahn-Woltjer model which gives a value of
the Local Group mass via the particular motions
of the two largest members,
the Milky Way and M31. Inclusion of dark energy in this model increases
the minimum mass estimate by
a factor of three compared to the "classical estimate". The increase is less
but still significant
for different ways of using the timing argument.
The second estimator is a modified virial
theorem which also demonstrates how dark energy can "hide" from detection a
part of the gravitating mass of the system. The third is a new zero-gravity
method which gives an upper limit to the group mass which we calculate 
with high precision HST observations. In combination,
the estimators lead to a robust and rather narrow range for a group's 
mass, M.  For the Local Group, $ 3.2 < M < 3.7 \times 10^{12} 
M_{\odot}$. Our result
agrees well with the Millennium Simulation based on the
$\Lambda$CDM cosmology.

\end{abstract}


\section{Introduction}

It has long been known that galaxy groups and clusters would fly
apart unless they were held together by the gravitational pull of
much more material than what is actually seen. The rapid progress
in dark matter studies stems from optical surveys of large areas
and high redshifts, CMB fluctuation measurements, sharp X-ray
images, gravitational lensing measurements, etc. Concurrently, 
conclusive observational indications have 
accumulated that dark matter and baryons contribute not more than
30\% to the mass/energy content of the universe. The rest 70\%
is composed of a much more mysterious dark energy (Riess et al.
1998, Perlmutter et al. 1999). Dark energy does not cluster, and
its simplest mathematical formulation goes back to 1917 when
Einstein introduced the cosmological constant $\Lambda$. This
interpretation is assumed in the currently standard $\Lambda$CDM
cosmology. A positive $\Lambda$ corresponds to a positive constant
energy density in all entire space. Such a medium with
negative pressure ($\rho_v > 0$, $p_v < 0$) is characterized by
the equation of state $\rho_v
= - p_v$ ($c =1$), like that of vacuum. According to Einstein's
equations, gravity depends on pressure as well as density: the
effective gravitating density $\rho_{\rm eff} = \rho + 3 p$  is
negative for a vacuum ($= -2 \rho_v$), leading to a
repulsion, or antigravity. The dark energy antigravity revealed
itself first in the accelerating cosmological expansion.

In this paper, we follow the $\Lambda$CDM cosmology and assume
that the dark energy (DE) is a perfect fluid with a constant density
every where and in any reference frame. From most recent CMB
studies (Spergel et al. 2006), $\rho_v = (0.72 \pm 0.03) \times
10^{-29}$ g cm$^{-3}$. We focus on relatively small spatial
scales, on the scale of groups of galaxies. We
show that galaxies and their systems "lose" a part of their
gravity due to the antigravity of the dark energy in
their volumes. The dynamical methods of the mass estimation of
dark matter (and baryons) should take into account this
"lost-gravity" effect -- otherwise the mass would systematically
be underestimated.

\section{Modified Kahn-Woltjer (MKW) estimator}

Fifty years ago, Kahn \& Woltjer (1959, KW) used simple linear two
body dynamics to describe the relative motion of the Milky Way and
M31 galaxies. The motion of the galaxies was described (in the
reference frame of the binary's center of mass) by the equation of
motion
\be \ddot D(t) = - GM/D^2, \ee where $D$ is the distance between
the galaxies and $M$ is the total mass $m_1+m_2$ of the binary. In
the well-known first integral
\be \frac{1}{2} \dot D(t)^2 = GM/D  + E_0, \ee \noindent
the constant total mechanical energy of the binary (per
unit reduced mass) is negative for a gravitationally
bound system,  $E_0 < 0$. This fact and Eq.2 with the
observed values $D = 0.7$ Mpc and $\dot D = -120$ km/s lead to an
absolute lower limit for the estimated binary mass: $M > 1 \times
10^{12} M_{\odot}$. 

The "timing argument" KW calculation integrates in reverse the 
equation of motion (1) from the above present-day separation and 
radial velocity requiring that the initial expansion of the two 
galaxies from one another started a time in the past equal to the age 
of the universe (13.7 billion years).  Thus
we calculate a mass for the pair of $M \simeq 4.0 
\times 10^{12} M_{\odot}$.
See Binney \& Tremaine (1987) for an analytic discussion (for the 10 
to 20 billion year age range suspected at that time).

With a minimal modification of the original KW method, including
the dark energy background $\rho_v$, the equation of motion
and its first integral become:
\be \ddot D(t) = - GM/D^2 + G \frac{8\pi}{3} \rho_{v}D, \ee
\be \frac{1}{2} \dot D(t)^2 = GM/D + G \frac{4\pi}{3} \rho_{v}D^2
+ E_0. \ee \noindent

Now the total energy for a bound system embedded in the DE
background is
\be E_0 < - \frac{3}{2} G M^{2/3} (\frac{8\pi}{3} \rho_v)^{1/3},
\ee \noindent as may be easily seen from the gravitational
potential $U = - GM/D - \frac{4\pi}{3} G \rho_{v}D^2$ in the right
side of Eq.4.  With the same data as above for current separation and 
the
relative velocity of the galaxies, Eqs. 4,5 lead to a new absolute
lower  mass limit: $M > 3.2 \times 10^{12} M_{\odot}$, or 3 times
the absolute lower limit of the original statement of the problem.

Again the MKW equations can be integrated back to the origin of the 
universe including the dark energy value given earlier.  The same data lead 
to a larger mass: $M = 4.5 \times
10^{12} M_{\odot}$.
 
In fact, the KW model can hardly be extrapolated to the origin of the universe,
while it seems realistic for the stage of the binary collapse. In Fig. 1
we show the calculated mass as a function of the time back to the
maximal separation of the Milky Way and M31; the lower and upper curves show the results
without and with dark energy, respectively. In the $\Lambda$CDM
cosmology gravitational instability is terminated in the linear
regime about 7 Gyr ago with the start of the DE dominated epoch
(e.g. Chernin et al. 2003). Taking this to be the minimal collapse time
yields the maximal mass of the binary, and this is $3.3 \times
10^{12} M_{\odot}$ without dark energy and $4.1 \times
10^{12} M_{\odot}$ with dark energy.

These results demonstrate clearly the effect of the lost gravity
in gravitationally bound systems embedded in the dark energy
background. The relative motion of the bodies is controlled by the
gravity (of dark matter and baryons) which is partly
counterbalanced by the DE antigravity. Consequently, the
mass estimate must be corrected: the
mass of the Local Group given by the modified Kahn-Woltjer (MKW)
estimator is significantly larger than that obtained
without inclusion of dark energy.
Here we have taken advantage of our position within the Local Group,
the well known distance of M31 and the measured approach speed.
We thus should modify conventional ways to estimate
the masses of groups of galaxies in general.

\section{Modified virial theorem (MVT)}

Conventional virial mass estimators use the relation between the
mean total kinetic $<K>$ and potential $<U>$ energies of a
quasi-stationary gravitationally bound many-body system: $<K> =
1/2 |<U>|$. The presence of dark energy modifies this relation.
The total potential energy includes now not only the sum $U_1$ of
the mutual potential energies of its member particles, but also
the sum $U_2$ of the potential energy of the same particles in the
force field of dark energy:
\be U_1 = - \frac{1}{2}\sum \frac{G m_i m_j}{|\textbf{r}_i -
\textbf{r}_j|}, \; U_2 = - \frac{4\pi \rho_v}{3}\sum m_i
\textbf{r}_i^2 \ee \noindent Here $\textbf{r}_i$ is the
radius-vector of a particle in the frame of the system's
barycenter; the summation in $U_1$ is over all particle pairs ($i
\neq j$). The major contribution to the sum is from dark matter
particles whatever their individual masses may be.

A hint to the structure of $U_2$ may be seen, e.g., from the
second item in the right side of Eq.4; the summation in $U_2$ is
over all the particles. Dark energy comes to the virial theorem
via an extra contribution to the potential energy of the system.

A link to the kinetic energy is provided by the equation of motion
of an individual particle:
\be m_i \ddot \textbf{r}_i = - \frac{\partial U}{\partial
\textbf{r}_i} = - \frac{\partial U_1}{\partial \textbf{r}_i} +
\frac{8\pi}{3}\rho_v m_i \textbf{r}_i^2 \frac{\textbf{r}_i}{r_i}.
\ee
Averaging over time and using the Euler theorem on homogeneous
functions (applied separately to the two functions that come from
the two terms in the right side of Eq.7), we find
\be <K> = - \frac{1}{2} <U_1> + <U_2>. \ee This is the new virial
relation adapted to the universe with the dark energy background.

Eq.8 may be rewritten in terms of the total mass $M$, a
characteristic velocity $\bar V$ and characteristic sizes $\bar
R_1, \bar R_2$:
\be M = \frac{\bar V^2 \bar R_1}{G} + \frac{8\pi}{3}\rho_v \bar
R_2^3.\ee
or in convenient units:
\be \frac{M}{M_{\odot}} = 2.3 \times 10^8 (\frac{\bar V}{\rm
km/s})^2(\frac{\bar R_1}{\rm Mpc})
 + 0.9 \times 10^{12}(\frac{\bar R_2}{\rm Mpc})^3
\ee
This new mass estimator -- the modified virial theorem (MVT)
estimator -- includes an additional positive term which is equal
to the absolute value of the effective (anti)gravitating mass of
dark energy contained in the spherical volume of the radius $R_2$:
$M_{eff} = - \frac{8\pi}{3}\rho_v \bar R_2^3$. This term gives a
quantitative measure of the lost-gravity effect.

One can get a physical sense of Eqs.9,10 by using representative values 
of the
characteristic size $\bar R_1, \bar R_2$ and velocity $\bar V$.
In the simplest example of only one body orbiting a
gravitating mass $M$, the size $\bar R_1 = \bar R_2$ is the
 radius of the orbit, the velocity $\bar V$ is the
orbital velocity. In this case, the total effective gravitating
mass within the orbit is the gravitating mass $M$ plus the
antigravitating (negative) mass $M_{eff}$. A similar
identification of the quantities is obvious as well when the dark
mass of a galaxy is derived from its rotation curve. In both
cases, the estimated mass is larger than that in conventional
estimates: the additional mass is $|M_{eff}|$.

Generally, the characteristic sizes and velocities for groups and
clusters need more sophisticated analysis - even in the absence of
dark energy (e.g. Peebles 1971).
Here we merely point out the expected significance of this effect
in small systems like the Local Group.
 The first (conventional) term in
Eq.9 is estimated for the Local Group as $2.3 \times 10^{12}
M_{\odot}$ (van den Bergh 1999) or  $1.9 \times 10^{12} M_{\odot}$
(Karachentsev 2007, Karachentsev et al. 2009). Then the second
term in Eq.9 increases the total mass by 30 to 50\%, if the
total size $\bar R_2$ is about 1 Mpc.
Such an effect is compatible with the other methods here discussed.

 Assuming $\bar R_1 = \bar
R_2$, one may see that the relative contribution of the dark
energy scales as the crossing time squared $(\bar R/\bar V)^2$. If
so, the lost-gravity effect is 10-30 times larger in a group like
the Local Group than in a rich cluster like the Coma cluster.

\section{Zero-gravity surface (ZGS) estimator }

As it was mentioned elsewhere (Chernin 2001, 2008; Byrd, Chernin
\& Valtonen 2007; Teerikorpi et al. 2008), an isolated
gravitationally bound galaxy group can exist on the dark energy
background, only if gravity dominates over antigravity in the
whole volume of the system. In a simple model, a group may be
represented by a spherical mass $M$ of dark matter and baryons,
embedded in the uniform dark energy background. Out of the mass at
a distance $R$ from the mass center, the gravity force is given by
Newton's inverse square law in the reference frame related to the
group barycenter. The antigravity force produced by the DE density
$\rho_v$ is given by Einstein's linear law (e.g. Chernin 2001,
2008, Chernin et al. 2006) in the same reference frame:
\be F_{\rm N} = - GM/R^2, \,\,\, F_{\rm E}  = + \frac{8
\pi}{3} G
 \rho_v R. \ee
Gravity and antigravity are exactly balanced at the zero-gravity
surface of the radius $R = R_{v}$ (Chernin et al. 2000; Chernin
2001; Baryshev et al. 2001, Dolgachev et al. 2003):
\be R_{v} = (\frac{3 M}{8\pi \rho_{v}})^{1/3}, \;\;\; R_v > R_0,
\ee \noindent where $R_0$ is the radius of the group. Antigravity
dominates at $R > R_v$, and gravity is stronger than antigravity
at $R<R_v$. The lost-gravity effect is one hundred percent
value on the zero-gravity surface: the total effective gravitating
mass of the non-vacuum matter and vacuum energy contained within
the surface is zero on the surface. No finite bound orbits are
possible on the surface and further at distances $R > R_v$.

Since the zero-gravity radius $R_v$ depends on the gravitating mass of the
group alone (for a fixed $\rho_v$), this mass may be found, if the
zero-gravity radius is known from observations:
\be M = {\frac{8\pi}{3}} \rho_v R_v^{3} \simeq 0.9 \times 10^{12}
[R_v/({\rm Mpc})]^{3} M_{\odot}. \ee \noindent

The position of the zero-gravity surface can indeed be detected.
The most complete high-precision data
come from systematic observations of galaxies up to the distance 3
Mpc performed recently with the Hubble Space Telescope
(Karachentsev 2005, 2007; Karachentsev et al. 2002, 2003, 2009).
About 60 galaxies with good distances (8-10\% accuracy) and
velocities (5-10 km/s accuracy) are observed in the volume. Nearly
half of them join the Milky Way and M31 galaxies forming the Local
Group. In the velocity-distance diagram they occupy
the distance range of about 1 Mpc;
their radial velocities (relative to the group barycenter)
are in the interval from -150 to +150 km/s. The other half of the
observed galaxies (all of them are dwarfs) move away from the group
and have only positive velocities in the diagram. It is important
that their minimal distances are 1.1-1.6 Mpc
from the group center (Karachentsev 2005, Chernin et al. 2007,
Karachentsev et al. 2009). The flow of receding galaxies is rather
regular: it follows mainly the linear velocity-distance relation,
and its velocity dispersion is only about 25-30 km/s.

This radical difference in the phase-space structure of the group
and the outflow around it is a most remarkable feature.
 We argue (Chernin et al. 2000, 2007, Chernin 2001,
Teerikorpi et al. 2008) that the physics behind this feature might
be due to the interplay between the gravity of the group and
antigravity of the dark energy background, so that the group size
$R_0$ is less than $R_v$ and the flow starts at the distances $R
> R_v$. Therefore the zero-gravity surface is located somewhere
in the gap between the group and the outflow, i.e. in the distance
interval from 1.1 to 1.6 Mpc.

The double inequality $1.1 < R_v < 1.6$ Mpc and Eq.13 lead to the
lower and upper limits for the non-vacuum mass of the group:
\be 1.2 < M < 3.7 \times 10^{12} M_{\odot}. \ee

This result is compatible with the
estimations obtained in Secs.2,3. The ZGS-estimator uses the
overall structure of the group as a whole, while the MKW- and
MVT-estimators deal with the internal dynamics of the group. The
new  ZGS-estimator is possible only due to the existence of dark energy 
in the universe.

\section{Conclusions and discussion}

The local volume proves to be a prospective arena for the study
of dark matter and dark energy (e.g. Chernin 2001, 2008;
Byrd, Chernin \& Valtonen 2007; Niemi et al. 2007;
Teerikorpi et al. 2008). We have demonstrated above that,
surprisingly enough, by taking the dark energy background into account
enables one to learn more on the mass of dark
matter in galaxies and their systems.

To summarize the results:

1. The minimally modified Kahn-Woltier method leads to a new
absolute minimum of the Local Group mass: $M > 3.2 \times 10^{12}
M_{\odot}$, and it is 3 times larger than the similar mass minimum
obtained via the original KW method.  The back integration 
leads to a mass of the M31, Milky Way pair of $4.5 \times 10^{12}
M_{\odot}$, significantly greater than the original method.

2. The modified virial theorem has the form: $<K> = -
{\frac{1}{2}} <U_1> + <U_2>$. It accounts for the potential energy
$U_2$ which is due to the antigravity force of dark energy. The
theorem gives a clear measure of the lost-gravity effect.

3. The Local Group is located in the area of its self-gravity
domination which makes it gravitationally bound, while the outflow
around the system develops in the area where no bound orbits are
possible. The zero-gravity surface separates these two
areas, and its location identified with the observed
velocity-distance diagram leads to the absolute upper limit of the
group mass:  $M < 3.7 \times 10^{12} M_{\odot}$.

Some additional remarks:

1. The linear binary model for the relative motion of the Milky
Way and M31 galaxies (Sec.2) is an idealization. The real dynamics
of the Local Group must be more complex (see the study by Valtonen
et al. 1993). In particular, a transverse velocity might increase
the mass value, but not more than by 10-15\%. The earlier
dynamical history of the binary (not even mentioning its formation)
can hardly be described by this model. This process needs more
studies, and cosmological N-body simulations are a
real tool to clarify the matter. The most recent Millennium
Simulation shows a mass range $(2-5) \times 10^{12} M_{\odot}$ for
galaxy groups similar to the Local Group (Li and White 2008); it
seems instructive that the interval found in Secs.2,4 is in a good
agreement with the simulation result.

2. A study of the lost-gravity effect in groups in general is
complicated by the fact that in existing group catalogues, the
virial masses tend to be overestimated for other reasons (Niemi et
al. 2007); we will discuss this in detail elsewhere. We will also
reexamine the large-scale mean dark matter and baryonic densities
and study the related problem of the dark mass deficit in the
local volume (Karachentsev et al. 2009).

3. The gravity potential of a galaxy group like the Local Group
which contains a dominant binary of giant galaxies is obviously
non-static and non-spherically symmetrical - contrary to the model
of Sec.4. However non-sphericity and time-dependence are not
significant when one considers relatively large distances from
the group barycenter. Calculations show that the deviations of the
zero-gravity surface from sphericity 
now and during the binary collapse are not
larger than about 10\% in radius (Dolgachev et al. 2003; Chernin et al. 2004).
The deviations are even less in the area of the outflow around the
group.

4. It is worthwhile to compare the zero-gravity method of Sec.4
with the classical method of the zero-velocity distance $R_{\rm
0}$ predicted by Lema\^{i}tre-Tolman solution without or with dark
energy (e.g. Peirani \& de Freitas Pacheco 2006, 2008,
Karachentsev et al. 2009). In that approach the equation of motion
for mass shells surrounding the central mass $M$ is solved by
integrating from a moment near the Big Bang. The resulting {\it
present} velocity-distance relation gives the turn-around radius
$R_{\rm 0}$, to be compared with the observations. The mass turns
out to be proportional to $h^2R_{\rm 0}^3$ and the coefficient of
proportionality is different for the solutions with or without the
cosmological constant, so that for a given $R_{\rm 0}$ the
inferred mass is again larger when dark energy is included (by
about 30\%; Peirani \& de Freitas Pacheco 2006). In contrast, our
work on the early dynamics of the Local Group suggests
that there is a finite time interval after the Big Bang
where the outflow of dwarf galaxies originates 
(e.g. Chernin et al. 2004, 2007; Byrd, Chernin \& Valtonen 2007). In
this case the present zero-velocity distance will not be uniquely
predicted, while, on the contrary, the zero-gravity distance has
its near-constant value depending just on $M$ and $\rho_v$.

Peirani \& de Freitas Pacheco (2008) did not use directly $R_{\rm
0}$, but fitted their LT-predictions through velocity-distance
data points around the Local Group and some nearby groups. They
did not get different mass estimates with or without the
cosmological constant. A possible reason may be the assumption of
the simultaneous origin of the outflow near the cosmological
singularity, and their
value $2.4 \times 10^{12} M_{\odot}$ should be regarded as a lower limit.

To conclude, the mass determination for galaxies and their systems
is still a hard observational problem - even for the nearest well
studied objects. The task needs also a reliable theory background.
The lost-gravity effect is a new significant element of the
underlying physics. It is included in three different and
self-consistent mass estimators
introduced above. Their results are compatible with each other,
and in combination they give a robust and rather
narrow range of the Local Group non-vacuum mass: $ 3.2 < M < 3.7
\times 10^{12} M_{\odot}$. Even for the lower limit (from
the MKW binary energy model), the individual dark halos of the Milky Way
and M31 galaxies contain together at least 90\% of the total dark
matter of the group, and the rest might be distributed in
the outer common halo of the group around its dominated binary.

\acknowledgements We thank A. Cherepashchuk, Yu. Efremov, A.
Silbergleit and A. Zasov for discussions. AC appreciates a partial
support from RFBR under the grant 09-02-00418.

\newpage

\begin{figure*}
\epsfig{file=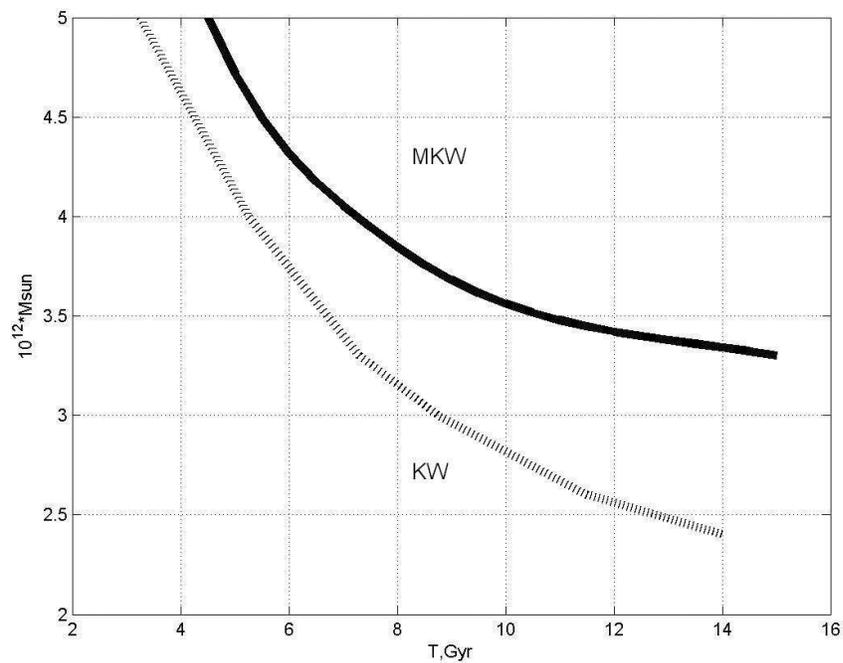, angle=0, width=13cm}
\caption{
This diagram shows the difference between the Local Group mass predictions from
the classical Kahn-Woltjer estimator (KW) and its modified form here
introduced (MKW). The x-axis gives the look-back time of maximum separation and
the y-axis gives the calculated total mass of the Milky Way \& M31 binary.}
\label{fig1}
\end{figure*}

\end{document}